\documentclass[sigconf]{acmart}
\acmConference{}{}{} 
\acmBooktitle{}
\acmPrice{}
\acmDOI{}
\acmISBN{}
\renewcommand\footnotetextcopyrightpermission[1]{}
\usepackage{tikz}
\usepackage{amsmath}
\usetikzlibrary{shapes, arrows, positioning}
\usepackage{graphicx}
\usepackage[colorinlistoftodos]{todonotes}
\usepackage{comment}
\usepackage{subcaption}
\usepackage{graphicx}

\definecolor{cinnabar}{rgb}{0.89, 0.26, 0.2}

\AtBeginDocument{%
  }

\begin{document}

\title{How Memory-Safe is IoT? Assessing the Impact of Memory-Protection Solutions for Securing Wireless Gateways}

\author{Vadim Safronov}
\email{vadim.safronov@cs.ox.ac.uk}
\affiliation{%
  \institution{University of Oxford}
  \city{Oxford}
  \country{United Kingdom}
}

\author{Ionut Bostan}
\email{ionut@nquiringminds.com}
\affiliation{%
  \institution{NquiringMinds}
  \city{Southampton}
  \country{United Kingdom}}

\author{Nicholas Allott}
\email{nick@nquiringminds.com}
\affiliation{%
  \institution{NquiringMinds}
  \city{Southampton}
  \country{United Kingdom}
  }

\author{Andrew Martin}
\email{andrew.martin@cs.ox.ac.uk}
\affiliation{%
  \institution{University of Oxford}
  \city{Oxford}
  \country{United Kingdom}
}

\begin{abstract}
The rapid development of the Internet of Things~(IoT) has enabled novel user-centred applications, including many in safety-critical areas such as healthcare, smart environment security, and emergency response systems. The diversity in IoT manufacturers, standards, and devices creates a combinatorial explosion of such deployment scenarios, leading to increased security and safety threats due to the difficulty of managing such heterogeneity. In almost every IoT deployment, wireless gateways are crucial for interconnecting IoT devices and providing services, yet they are vulnerable to external threats and serve as key entry points for large-scale IoT attacks. Memory-based vulnerabilities are among the most serious threats in software, with no universal solution yet available. Legacy memory protection mechanisms, such as canaries, RELRO, NX, and Fortify, have enhanced memory safety but remain insufficient for comprehensive protection. Emerging technologies like ARM-MTE, CHERI, and Rust are based on more universal and robust Secure-by-Design (SbD) memory safety principles, yet each entails different trade-offs in hardware or code modifications. Given the challenges of balancing security levels with associated overheads in IoT systems, this paper explores the impact of memory safety on the IoT domain through an empirical large-scale analysis of memory-related vulnerabilities in modern wireless gateways. Our results show that memory vulnerabilities constitute the majority of IoT gateway threats, underscoring the necessity for SbD solutions, with the choice of memory-protection technology depending on specific use cases and associated overheads.
\end{abstract}

\renewcommand{\shortauthors}{Safronov et al.}

\ccsdesc[500]{Security and privacy~Systems security}
\ccsdesc[500]{Networks~Wireless access points, base stations and infrastructure}

\keywords{Security, Memory Safety, Internet of Things, Firmware Analysis}


\maketitle

\section{Introduction}
The rapid rise of IoT devices and applications is revolutionising various industries. Currently, there are billions of deployed IoT devices, with billions more expected in the coming years~\cite{iot2024connected}. To provide users and businesses with a wide array of smart features, such as adaptive climate control, smart parking, traffic safety, healthcare, and smart environmental safety, IoT devices are manufactured globally, transmit data worldwide, receive over-the-air updates from various locations, and utilise a myriad of third-party software libraries. This uncontrolled heterogeneity in the production of IoT devices, their communication patterns, and applications may result in significant safety, security and privacy threats on a global scale. Even a small vulnerability in one component of this complex IoT ecosystem~\cite{conti_iot, copos_iot} could significantly disrupt the entire security and safety infrastructure of large organisations, including cyber-critical systems~\cite{Karale_iot, Swessi_iot}, such as healthcare or emergency response systems. 

A large number of cyberthreats can be triggered by memory-based vulnerabilities~\cite{memory_attack}, resulting in broad and intricate attack chains. For example, a single compromised router can become the entry point for attacking IoT devices, which subsequently compromises entire networks, escalating to widespread cyber threats. Notable incidents highlight the urgency of this issue: the Dyn DNS attack, where a multitude of IoT devices were used to launch a massive Distributed Denial of Service (DDoS) attacks, the SolarWinds breach, and Log4j incidents all exemplify the wide systemic consequences of vulnerabilities (however they arise). 

Although risk management has helped to stabilise many contexts, incident responses, and disaster containment plans, it is increasingly clear that step-changes are needed in order to eliminate whole classes of vulnerabilities.  Doing so offers immense potential for improved security resilience: failure to address these fundamentals will leave cyber space vulnerable to unpredictable systemic failures indefinitely, posing risks not just to individual systems but to large ecosystems, the wider economy, and public health.

Millions of new IoT devices emerge annually, potentially introducing novel, unknown security threats, partly due to the lack of centralised control and standardisation. Moreover, this trend highlights the continuous introduction of fresh vulnerabilities, both in new systems and in the patches intended to fix old ones. This procession of endlessly-broken interconnected IoT devices, networks and systems pushes us to the emerging paradigm of Security by Design~(SbD).  Whilst system architects have long understood --- or been enjoined to take account of --- the security of their systems and the privacy of the data they process, it has not always been a priority.  Improved designs --- and better implementation primitives --- help to eliminate large classes of vulnerabilities. The UK Government’s ‘Digital Security by Design’ initiative is an example of such an approach, promoting the CHERI architecture with a promise of low-cost memory protection, claimed to eliminate at a stroke around 70\% of the vulnerabilities present in a wide range of systems, according to analysis from Microsoft Security Response Centre~\cite{miller2019trends}.

Along with the CHERI architecture, other promising SbD approaches for memory safety include memory-safe languages like Rust~\cite{matsakis2014rust} and memory-tagging techniques such as ARM-MTE~\cite{arm}. Each approach involves trade-offs between security guarantees and associated overheads. Rust and CHERI offer deterministic memory protection, capable of defending against known and future memory-safety threats, but require hardware modifications for CHERI and significant code rewrites for Rust during the porting of memory-unsafe software. In contrast, memory-tagging techniques like ARM-MTE have less security robustness due to the probabilistic nature of the tagging approach, yet they require fewer hardware and code changes, making them easier to implement in modern processors and to adapt to existing software.

\subsection{Research questions}

Motivated by the urgent need for a thorough investigation into the cyber threats prevalent in IoT and the role of memory protection in this domain, our study maps out and classifies cyber threats within wireless gateway firmware, exploring the impact of memory-protection technologies on their mitigation. We conducted an empirical large-scale vulnerability analysis based on the examination of 6,335 firmware images of modern wireless gateways. The focus on wireless gateways stems from the fact that routers serve as sweet spots for attacks originating both from inside and outside IoT networks, with the potential to propagate across entire network infrastructures, particularly as they integrate IoT networks with safety-critical systems like building security, traffic management, and healthcare. Specifically, our research addresses the following questions:

\begin{itemize}
    \item \textbf{RQ1:}What is the ratio and classification of memory safety threats compared to all vulnerabilities in modern wireless gateway firmware? (\textsection \ref{analysis})
    \item \textbf{RQ2:} What is the quantitative impact of memory protection on wireless gateways, and how effectively can SbD solutions guarantee it in terms of vulnerability coverage, transfer overheads, and future-proof potential? (\textsection \ref{impact})
\end{itemize}

\subsection{Contributions}
The key contributions of our research are as follows.
\begin{itemize}
    \item We empirically analysed existing vulnerabilities in wireless gateway firmware~(\textsection \ref{analysis}). From a total of 6,335 router firmware images, we sampled and analysed 502 firmware binary samples, identifying 17,341 occurrences of common vulnerabilities and exposures (CVEs). We quantitatively categorised the identified vulnerabilities by memory relevance and severity level, providing insights into the ratio and types of memory safety threats in relation to all identified threats.
    \item Based on the vulnerability analysis, we quantitatively measured the impact of SbD memory protection~(\textsection \ref{impact}). Our findings show that deploying SbD solutions in wireless gateways can eliminate 74\% of known CVEs, increasing the security of average router firmware by a factor of 3.8. This underscores the necessity of SbD solutions for enhancing the security of IoT, particularly in safety-critical applications. 
    \item Based on the SbD impact evaluation, this work discusses promising SbD solutions, highlighting their strengths and weaknesses in terms of security, performance, and economic perspectives, suggesting future research directions~(\textsection \ref{impact}).
\end{itemize}

\section{Related work}

\textbf{IoT firmware analysis.} Recent advancements in firmware analysis for IoT vulnerability detection include several notable studies. Li et al.\cite{IoT_1} survey current methods for static, dynamic, and hybrid analysis, highlighting challenges in detecting zero-day vulnerabilities in Linux-based IoT devices. Zhao et al.\cite{IoT_2} introduce FirmSec, a tool that identifies third-party components (TPCs) in firmware, revealing significant security issues, including outdated TPCs and regional disparities in vulnerability severity. Yu et al.\cite{IoT_3} examine over 10,000 Linux-based firmware images of embedded devices, finding low adoption rates of standard attack mitigations, indicating a rising security threat in IoT. Wu et al.\cite{IoT_4} present ChkUp, a tool for identifying vulnerabilities in firmware update processes, which led to the discovery of several previously unknown CVEs. While these works on firmware analysis enhance vulnerability detection, our contribution differs by focusing on the classification of vulnerabilities into memory-related categories and the assessment of the impact of memory protection mechanisms on IoT.

\textbf{Memory protection in Safety and Security by Design.} Safety-by-Design is a multidisciplinary concept related to fields such as engineering, healthcare, and environmental sciences, where proactive safety measures are integrated into design processes to prevent hazards and mitigate risks~\cite{safety-by-design}. In the IoT domain, especially safety-critical smart applications, this concept is closely related to Security by Design~(SbD), which helps prevent safety threats through robust, proactive security design practices. Memory safety is a critical SbD research area in securing modern computing systems, contributing to the overall SbD goals. Recent works on SbD approaches have focused on exploring memory-tagging and capability-based architectures for improving memory protection. While numerous SbD solutions have been proposed, some are outdated or remain largely theoretical, with CHERI~\cite{CHERI} and ARM-MTE~\cite{arm} emerging as two of the most promising technologies, each offering distinct trade-offs for memory safety by design. A comparative survey~\cite{tagged_memory_protection_survey} reviews hardware-based memory-protection methods, including CHERI and ARM-MTE, and highlights the challenges of adopting these and other SbD architectures, such as policy limitations, language ambiguity, memory performance overheads, increased silicon area, and higher power consumption. Xu et al.~\cite{infat_pointer} emphasise spatial memory safety through hardware-assisted approaches bazed on tagged pointers, noting the importance of subobject granularity. Na et al.~\cite{spectre_era_mitigations} explore memory corruption mitigations against speculative execution vulnerabilities, with CHERI and ARM-MTE included among the promising memory protection architectures. Memory-safe languages complement hardware protection, with Rust gaining attention for enforcing memory safety at compile time without compromising performance~\cite{matsakis2014rust}. RustBelt~\cite{jung2017rustbelt} verifies Rust's type system, ensuring reliability. 
As a memory-safe language, Rust shows potential for reducing security risks in widespread C/C++ codebases, while legacy tools for integrating memory safety with C/C++~\cite{serebryany2012addresssanitizer, clang2012memorysanitizer} offer only partial solutions for memory safety, with added runtime overhead and less comprehensive coverage. C++17/20 libraries, such as \texttt{std::span} and \texttt{std::unique\_ptr}, improve memory safety for specific use cases, such as bounds-checked access and pointer ownership management. However, these solutions are limited to their intended scenarios and do not fully address unknown or emerging memory safety threats by design.

\section{Methodology}
\label{method}

\subsection{Firmware dataset}
For measuring memory protection impact, we selected a firmware dataset used in recent CVE analysis of IoT firmware images~\cite{firmsec_paper}. From each vendor, we randomly selected 502 firmware binary samples from a total of 6,335 router software images of popular vendors. The distribution of the samples across the chosen vendors is as follows: TP-Link (138), D-Link (54), Phicomm (106), Trendnet (100), and OpenWrt (104).

\subsection{Firmware analysis}

For gateway firmware analysis, we employed our Software Bill of Materials Generation and Analysis Platform (SBOM-GAP)~\cite{SBOM-GAP}, an AI-based firmware analysis tool that enables rapid, large-scale vulnerability assessments based on Software Bill of Materials~(SBOM) principles. The concept of an SBOM, initially discussed in the manufacturing sector~\cite{articleBOM}, gained formal recognition in the United States in 2021~\cite{NTIA_SBOM, Biden2021Cybersecurity} as a promising approach to improving software transparency and security, with the European Union also expressing interest in adopting it~\cite{EU_SBOM_Work2023}.


SBOM-GAP generates and analyses SBOMs across a variety of software ecosystems, including binary files, container images, and projects with source code. Such a tool is crucial for securing IoT systems, as it provides in-depth analysis and classification of existing vulnerabilities through the systematic extraction and examination of software components. While SBOM-GAP is designed for a range of software testing and analysis objectives, applicable to more than just IoT systems, this paper focuses solely on the features relevant to its application in our study, covering firmware analysis of wireless network gateways. The SBOM-GAP pipeline consists of four core stages, described below.

\textbf{(1)~Binary extraction.} The SBOM-GAP tool embeds the Binwalk utility~\cite{binwalk} for extracting firmware binary images. Binwalk scans the binary for known file signatures and filesystems, such as SquashFS, JFFS2, or CramFS. Upon identifying a filesystem, it extracts and unpacks the data, providing access to the root filesystem. This process exposes the full directory structure and files within the firmware, which are essential for subsequent analysis of the core software components and configurations.

\textbf{(2)~SBOM generation from root filesystem.} SBOM-GAP utilises the Syft library~\cite{syft} to generate an SBOM from the extracted binary by scanning the root filesystem. This process identifies and lists all software packages, libraries, and dependencies, creating a detailed inventory of the available components. A typical SBOM contains a list of Common Platform Enumerations~(CPEs), which represents names of software components acccording to the NIST CPE naming convention. For example, version 0.9.3 of the OpenSSL library is encoded as \texttt{cpe:2.3:a:openssl:openssl:0.9.3:-::::::}. The SBOM is generated in CycloneDX JSON format~\cite{cyclonedx}, which is widely supported by most analysis tools and serves as the foundation for further vulnerability assessments.

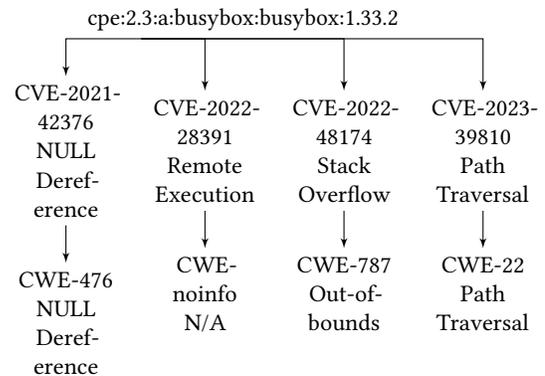
\begin{figure}[ht]
    \centering
    \begin{tikzpicture}[auto, node distance=0.5cm and 0.2cm,>=latex']
        \tikzset{block/.style={text width=4.5em, text centered, minimum height=1em}}

        \node [block] (cpe) {cpe:2.3:a:busybox:busybox:1.33.2};
        \node [block, below=of cpe, xshift=-1.0cm] (cve1) {CVE-2021-42376\\NULL Dereference};
        \node [block, below=of cve1] (cwe1) {CWE-476\\NULL Dereference};
        \node [block, right=of cve1] (cve2) {CVE-2022-28391\\Remote Execution};
        \node [block, below=of cve2] (cwe2) {CWE-noinfo\\N/A};
        \node [block, right=of cve2] (cve3) {CVE-2022-48174\\Stack Overflow};
        \node [block, below=of cve3] (cwe3) {CWE-787\\Out-of-bounds};
        \node [block, right=of cve3] (cve4) {CVE-2023-39810\\Path Traversal};
        \node [block, below=of cve4] (cwe4) {CWE-22\\Path Traversal};

        \draw [->] (cpe.south) -| (cve1.north);
        \draw [->] (cve1) -- (cwe1);
        \draw [->] (cpe.south) -| (cve2.north);
        \draw [->] (cve2) -- (cwe2);
        \draw [->] (cpe.south) -| (cve3.north);
        \draw [->] (cve3) -- (cwe3);
        \draw [->] (cpe.south) -| (cve4.north);
        \draw [->] (cve4) -- (cwe4);
    \end{tikzpicture}
    \caption{Relationship between CPE, CVE, and CWE for Busybox 1.33.2.}
    \label{fig:busybox-cpe-cve-cwe}
\end{figure}

\textbf{(3)~SBOM-based CVE/CWE extraction.} SBOM-GAP processes the generated SBOM to extract Common Vulnerabilities and Exposures~(CVEs) and Common Weakness Enumerations~(CWEs) associated with the identified CPEs. CVEs represent specific software vulnerabilities, while CWEs describe the underlying weaknesses that can give rise to these vulnerabilities. SBOM-GAP queries NIST and other databases to highlight the risks each component poses to the application. Figure~\ref{fig:busybox-cpe-cve-cwe} demonstrates CPE, CVE and CWE relationships based on the busybox CPE example.

\textbf{(4)~AI-based CVE/CWE classification into memory classes.} For rapid automated classification of CVEs and CWEs into memory-related categories, SBOM-GAP employs a GPT-4o-based OpenAI foundation model~\cite{openai_gpt4o} to classify vulnerabilities according to their memory-related classes. The process involves sending vulnerability descriptions to the GPT model using a structured prompt that instructs the model to classify the vulnerability as ``not-memory-related", ``spatial-memory-related", ``temporal-memory-related", or ``other-memory-related". The model then returns a JSON object that includes the reasoning and classification for each vulnerability. This LLM-based classification method was successfully tested and validated on manual classification on a sample of 177 CVEs and CWEs.
\section{Vulnerability analysis and classification}
\label{analysis}
To address RQ1, we conducted an empirical quantitative analysis of vulnerabilities present in gateway firmware, classifying them into different memory-related categories using the SBOM-GAP firmware analysis tool, according to the defined methodology. For each firmware image in the dataset, SBOM-GAP extracted the filesystem and generated an SBOM file in JSON CycloneDX format. Each SBOM was then parsed for distinct CPEs, and for each CPE, SBOM-GAP retrieved a list of associated CWEs and CVEs, along with their descriptions and Common Vulnerability Scoring System~(CVSS) scores. Since each CVE is related to a specific CWE, memory classification was primarily based on the CWE. This classification was further refined using the GPT-4o OpenAI model. If a CVE was not linked to any CWE, it was classified into a memory category based on its own description. Vulnerability occurrences were counted by aggregating all CVEs across dependencies within the firmware and across all firmware images. If the same CVE/CWE appeared in different components of a single firmware, each occurrence was counted separately, as each instance represents an independent potential threat.

\begin{figure}[h!]
    \centering
    \includegraphics[width=\linewidth]{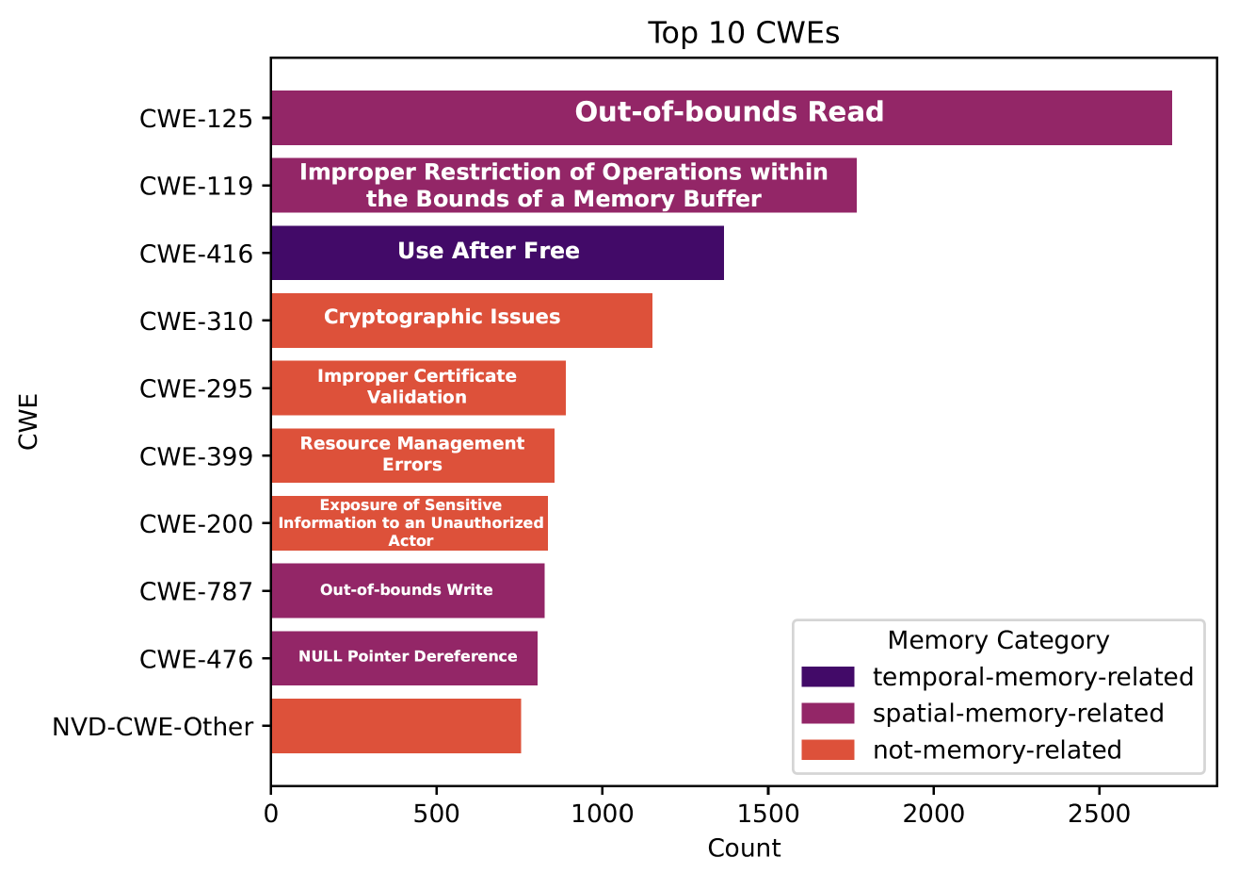}
    \caption{Top 10 CWEs across the analysed wireless gateway firmware. The "Count" axis shows the number of CWE occurrences across all libraries/dependencies found in the extracted firmware. The "CWE" axis represents CWE identifiers from the CWE MITRE database~\cite{cwe_mitre}.}
    \label{fig:top_10_cwes}
\end{figure}

\textbf{Most popular CWEs.} Figure~\ref{fig:top_10_cwes} shows the most widespread CWEs across the analysed gateway firmware. Although only half of the top 10 CWEs belong to memory-related categories, these account for approximately 70\% of all analysed CWEs, with spatial memory issues being the most prevalent. Notably, the top 3 CWEs are all memory-related: CWE-125 (Out-of-Bounds Read, spatial), CWE-119 (Improper Restriction of Operations within the Bounds of a Memory Buffer, spatial), and CWE-416 (Use After Free, temporal).

\begin{figure}[h!]
    \centering
    \includegraphics[width=\linewidth]{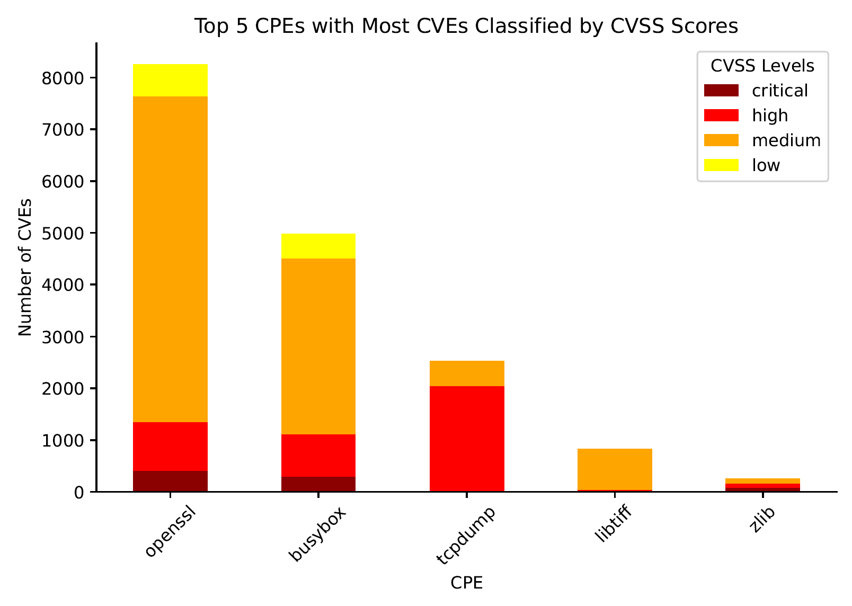}
    \caption{Top 5 CPEs across the analysed wireless gateway firmware images, classified by CVSS score. The "CPE" axis shows the CPE names of software libraries/dependencies found in the extracted firmware. The "Number of CVEs" axis shows the CVE counts for each CPE.}
    \label{fig:top_5_CPEs_CVSS}
\end{figure}

\begin{figure}[h!]
    \centering
    \includegraphics[width=\linewidth]{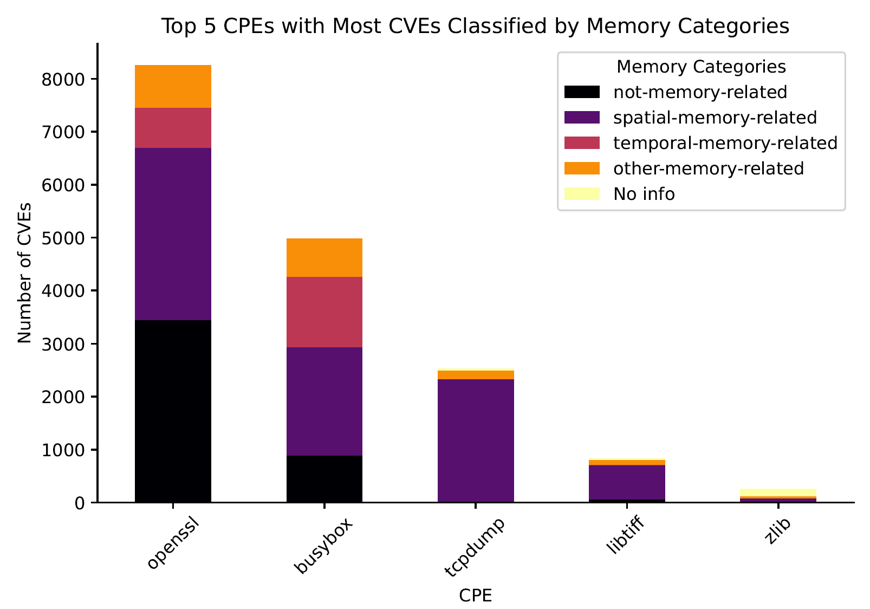}
    \caption{Top 5 CPEs across the analysed wireless gateway firmware images, classified by memory relevance. The "CPE" axis shows the CPE names of software libraries/dependencies found in the extracted firmware. The "Number of CVEs" axis shows the CVE counts for each CPE.}
    \label{fig:top_5_CPEs_memory}
\end{figure}

\textbf{CVE distributions.} Figures~\ref{fig:top_5_CPEs_CVSS} and~\ref{fig:top_5_CPEs_memory} represents CVE distributions according to the top 5 most popular third-party libraries and by their classifications into their severity~(CVSS score) and memory relevance. It can be seen that most found vulnerabilities are memory-related, with the majority distributed across OpenSSL, BusyBox, and tcpdump third-party software. Although, according to Figure~\ref{fig:top_5_CPEs_CVSS}, the majority of CVEs has medium CVSS score, it is evident that approximately 4500 CVEs have high and above criticality, meaning that on average, each wireless gateway image has 9 CVE occurences of such high/critical class. In terms of CVE distribution by memory classes~(Figure\ref{fig:top_5_CPEs_memory}), 57, 80, 100\% of vulnerabilities within the top 3 used libraries are memory-related, forming the major class of all other vulnerabilities. 

\textbf{CVE distributions.} Figures~\ref{fig:top_5_CPEs_CVSS} and~\ref{fig:top_5_CPEs_memory} present the distribution of CVEs according to the top 5 most popular third-party libraries, classified by their severity (CVSS) score and memory relevance. It is evident that most of the identified vulnerabilities are memory-related, with the majority found in OpenSSL, BusyBox, and tcpdump libraries. Although Figure~\ref{fig:top_5_CPEs_CVSS} shows that the majority of CVEs have a medium CVSS score, approximately 4,500 CVEs are classified as high or critical, indicating that, on average, each wireless gateway firmware has 9 occurrences of such high/critical vulnerabilities. In terms of CVE distribution by memory classes (Figure~\ref{fig:top_5_CPEs_memory}), 58.13\%, 82.13\%, and 100\% of vulnerabilities within OpenSSL, BusyBox, and tcpdump respectively are memory-related, making this the predominant category among all other vulnerabilities.

\section{Estimating memory safety impact and discussing SbD technologies}
\label{impact}

To address RQ2, which focuses on assessing the impact of memory protection and exploring promising Security by Design (SbD) technologies, we start by empirically evaluating the overall effectiveness of an ideal SbD memory protection solution. Following this, we examine the range of available SbD solutions, emphasising their respective trade-offs in terms of security, cost, and operational overhead.

\subsection{Memory protection impact} 

\begin{figure}[h!]
    \centering
    \includegraphics[width=\linewidth]{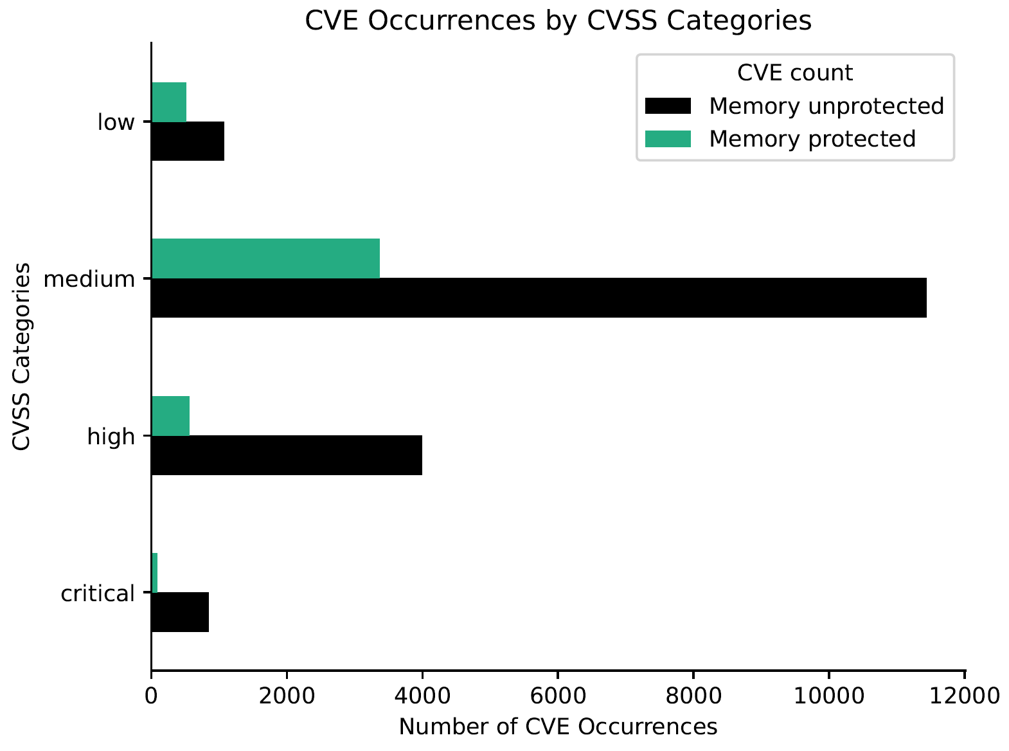}
    \caption{Impact of Secure-by-Design (SbD) memory safety on reducing CVE occurrences in wireless gateway firmware, categorised by CVSS scores. The results demonstrate the substantial potential of SbD memory safety solutions to reduce vulnerability frequency across all severity categories.}
    \label{fig:memory-protection}
\end{figure}

Figure~\ref{fig:memory-protection} illustrates the differences in the total number of CVE occurrences across the analysed wireless gateways with SbD memory protection employed compared to the existing state. The bar chart shows that critical/high CVEs are almost eliminated with memory protection in place, while medium CVEs are reduced by threefold. Overall, the total number of vulnerabilities decreases by approximately 3.5 times. These empirical numbers are based on an analysis of a sample of 502 router images from popular vendors, with the expectation that the difference will increase with further firmware analysis. On average, a wireless gateway would experience the following reduction in existing vulnerabilities: from 2 to approximately 0 critical CVEs, from 8 to 1.4 high CVEs, from 23.5 to 7.6 medium CVEs, and from 2 to 1 low CVEs.

\subsection{Comparing SbD memory-protection technologies}

Memory protection technologies are categorised into deterministic, probabilistic, and mixed SbD approaches. This analysis does not consider legacy techniques, as they offer only partial coverage of memory vulnerabilities. For instance, Stack Canaries and Fortify Source target issues such as stack and buffer overflows, while NX and RELRO provide broader protection but are not designed to address critical vulnerabilities like use-after-free, frequently observed in our firmware analysis. Consequently, these approaches are insufficient for delivering comprehensive memory safety in modern IoT systems.

\textbf{Deterministic SbD approaches.} Both CHERI~(capability-based protection) and Rust (a memory-safe programming language) offer comprehensive, multidimensional memory safety, deterministically covering memory-related CVEs. CHERI works by using hardware-enforced capabilities to restrict how memory can be accessed, ensuring pointers are within bounds and properly authorised. Rust, on the other hand, ensures memory safety at the language level through strict ownership and borrowing rules that prevent common vulnerabilities like buffer overflows and use-after-free. CHERI requires new processor architectures with low code modification overhead to leverage its capability features, whereas Rust necessitates a complete software rewrite, as most existing CPEs in wireless gateways are written in C.

\textbf{Probabilistic SbD approaches.} MMemory tagging technologies, such as ARM-MTE, provide multidimensional memory safety with minimal code porting, associating tags with memory addresses to detect illegal access, such as out-of-bounds or use-after-free errors. ARM-MTE, supported by ARMv8.5+, is attractive for its low integration cost. Microsoft's analysis~\cite{microsoft_arm-mte} suggests that the probabilistic nature of memory tagging may result in up to 6\% of cases where certain spatial and temporal memory-related CWEs are not fully mitigated. While this approach remains effective in most scenarios, it may be a consideration for safety-critical applications requiring the highest levels of reliability and security.

\textbf{Mixed SbD approaches.} Various mixed SbD strategies combine hardware and software to balance security and practicality. For safety-critical applications, businesses might adopt new hardware like CHERI for deterministic protection. The highest protection combines CHERI with Rust or other memory-safe languages, though this comes with significant costs, including new hardware and extensive code porting. In less critical scenarios, ARM-MTE offers a balanced solution with lower overheads and better software compatibility. Pairing Rust with ARM-MTE can enhance security and safety by providing deterministic software mitigation over a probabilistic hardware foundation.

\section{Conclusion}
The IoT positively transforming various industries, yet it has also introduced substantial safety and security risks due to its uncontrolled heterogeneity and lack of standardisation in place. Our research highlights that memory-based vulnerabilities is a critical and dominant threat in wireless gateways, which are frequent targets for IoT cyberattacks and thus can expose entire networks to large-scale risks, including potential disruptions in safety-critical areas such as healthcare, environmental safety, and security. Our large-scale analysis of wireless gateway firmware underscores the urgent need for robust memory protection strategies, especially in safety-critical IoT deployments, to mitigate these significant risks.

This work demonstrates that deploying Secure-by-Design (SbD) solutions significantly enhances the security of wireless gateways, increasing resilience to memory-safety vulnerabilities. Given that most code in wireless gateways is written in C, transitioning to CHERI or ARM-MTE is a more practical solution than rewriting it in a different memory-safe language. The selection of memory protection technology should be tailored to each deployment's specific needs, balancing security guarantees with overheads to effectively safeguard IoT environments in safety-critical applications.

\section*{Acknowledgements}
This work was supported by the Innovate UK-funded Secure Networking by Design (SNbD) project, grant number 10028034.

\bibliographystyle{ACM-Reference-Format}
\bibliography{references}

\end{document}